\documentclass[onecolumn,showpacs,aps,amssymb,floatfix,prd,amsmath,preprintnumbers]{revtex4}
\usepackage{epstopdf}
\usepackage{capt-of}
\usepackage{graphicx}  
\usepackage{dcolumn}   
\RequirePackage{graphicx}
\RequirePackage{float}
\RequirePackage{hyperref}
\RequirePackage{amsmath}
\RequirePackage{amssymb}
\RequirePackage{mathtools}

\begin{document}
\title{Stability analysis of anisotropic Bianchi type-I cosmological model
in teleparallel gravity }
\author{M. Koussour$^{1,\thanks{%
e-mail: pr.mouhssine@gmail.com}}$ and M. Bennai,$^{1,2,\thanks{%
e-mail: mdbennai@yahoo.fr}}$ \\
$^{1}${\small Quantum Physics and Magnetism Team, LPMC,}\\
{\small Faculty of Science Ben M'sik, Casablanca Hassan II University,
Morocco.}\\
$^{2}${\small Lab of High Energy Physics, Modeling and Simulations,}\\
{\small Faculty of Science, University Mohammed V-Agdal, Rabat, Morocco.}}
\begin{abstract} 
\textbf{Abstract:}In this work, we study a cosmological model of Bianchi type-I Universe in teleparallel gravity for a perfect fluid. To obtain the cosmological solution of the model, we assume that the deceleration parameter is a linear function of the Hubble parameter $H$ i.e. $q=-1+\beta H$ (where $\beta $ as a positive constant). Consequently, we get a model of our Universe, where it goes from the initial phase of deceleration to the current phase of acceleration. We have discussed some physical and geometric properties such as Hubble parameter, deceleration parameter, energy density, pressure, and equation of state (EoS) parameter and study their behavior graphically in terms of redshift and compare it with observational data such as Type Ia supernovae (SNIa). We also discussed the behavior of other parameters such as the Jerk parameter, Statefinder parameters and we tested the validity of the model by studying the stability analysis and energy conditions.
\end{abstract}
\maketitle
\section{Introduction}

Since Einstein \cite{ref1} published his theory of General Relativity (GR)
in 1916 until the end of the previous century, cosmologists believed that
the Universe was in a decelerated phase of expansion, due to Friedmann's
equations in the standard model of cosmology \cite{ref2}. But recently, a
group of theoretical and observational studies appeared in cosmology which
showed the opposite, that is to say, that our current Universe is in a phase
of accelerated expansion \cite{ref3, ref4, ref5, ref6, ref7}. This
contradiction between theories and observational data has led many
researchers to suggest other alternatives that agree between the two \cite%
{ref8}. The most famous is the idea that there is a new form of energy
called dark energy (DE) which is causing the accelerating expansion of the
Universe. According to the observational data, DE is characterized by
negative pressure or in some other way a negative EoS parameter $\omega $,
where $\omega =\frac{p}{\rho }$ with $\rho $ represents the energy density
of the Universe and $p$ represents the pressure. Many researchers recommend
another idea, namely that the accelerating expansion of the Universe could
be the result of modifications of gravity, and they called this class the
Modified Gravity Theories (MGT), see \cite{ref9, ref10, ref11, ref12, ref13,
ref14}. Another class combines the holographic principle with DE, such as 
\cite{ref15}. Many researchers around the world are striving to uncover the
causes of cosmic acceleration, but despite this, the question of cosmic
acceleration or DE remains a mystery in the scientific arena. Another simple
idea about the nature of DE in the context of GR was to add the cosmological
constant $\Lambda $\ that Einstein had introduced into his equations in
another context, but soon many problems arose, for example, it was a
difficulty on the theoretically predicted order of magnitude compared to
that of the observed vacuum energy \cite{ref16, ref17}.

Recently, a new type of study has appeared, which attempts to explain the
accelerated expansion of the Universe by assuming cosmological models in
various gravitational theories that contain the deceleration parameter or in
another way the scale factor which varies over time \cite{ref18, ref19,
ref20}. Indeed, this hypothesis is supported by observational data which
shows that the Universe has passed from the stage of early deceleration $%
(q>0)$ to the stage of current acceleration $(q<0)$.

Among all these modified gravity theories found in the literature, in this
work, we will focus on another approach to examine alternatives to GR which
is Teleparallel Gravity (TG) which uses the Weitzenbock connection in place
of the Levi--Civita connection and therefore does not have curvature but has
a torsion which is responsible for the acceleration of the Universe. Some
studies on this subject have gone so far as to replace scalar torsion $T$\
with a generalized function $f(T)$ for the latter \cite{ref21}. For example,
Bhoyar et al. study Bianchi type-I space-time for the linear and quadratic
form of $f(T)$ gravity with a hybrid expansion law for scale factor \cite%
{ref22}. Holographic DE has also been discussed in this context by Shaikh et
al. using the power and exponential laws \cite{ref23}. In another work,
Shaikh studies DE in a teleparallel gravity framework using the hybrid
expansion law for scale factor with the thermodynamic aspects of the model 
\cite{ref24}. See also other works in this context \cite{ref25, ref26,
ref27, ref28, ref29}. \\
Several studies support the idea that the geometry of
the Universe at the end of the inflationary era is homogeneous and isotropic 
\cite{ref30}, so that FLRW models play an important role in this period.
However, defects in the cosmic microwave background (CMB) due to quantum
fluctuations of the inflation period confirm the existence of an anisotropic
phase which then transformed into an isotropic phase. Recently, with the
advent of Planck observational data \cite{ref31}, Bianchi cosmological
models describing the anisotropic Universe have attracted the attention of
many authors. In the literature, there are several types of anisotropic and
inhomogeneous Bianchi space-times. The Bianchi type-I space-time is the
mathematically simplest case that describes an anisotropic and homogeneous
Universe. Also defined as a direct generalization of the FLRW Universe with
a different scale factor in each spatial direction. Several Bianchi type-I
cosmological models have been studied in various contexts by many
researchers. Recently, Hossienkhani et al. studied the spatially homogeneous
and anisotropic Bianchi type-I Universe with the interacting holographic and
new agegraphic scalar fields models of DE \cite{ref32}. The dynamical
evolution of an $f(R)$ model of gravity in a viscous and anisotropic
background given by Bianchi type-I space-time is discussed by Saaidi et al. 
\cite{ref33}. Also, the anisotropy effects on Baryogenesis in $f(R)$ gravity
are examined in Ref. \cite{ref34}.

In this paper, motivated by the above works we study a Bianchi type-I
cosmological model under teleparallel gravity $f\left( T\right) =T$ with the
perfect fluid-like matter. We assume that the deceleration parameter varies
with time as a linear function of the Hubble parameter i.e. $q=-1+\beta H$
(with $\beta $ as a positive constant and $H$ as Hubble parameter). Next, we
find solutions to the field equations and discuss some of the geometric and
physical properties of the model and compare them with the observational
data. The article is organized as follows: In Sect. 2, we gave a brief
description of the mathematical formalism of $f(T)$ gravity and its field
equation. In Sect. 3, we describe the metric of the Universe and the basic
equations of our model. We find solutions to the field equations by assuming
that the deceleration parameter is a linear function of the Hubble
parameter, and we discuss the behavior of each of the last two parameters
and compare them to the current values of the observational data in Sect. 4.
Sect. 5 is devoted to the physical and geometric properties of the model. In
the last section, the main results of the model are discussed.

\section{$f(T)$ gravity formalism}

As usual, in this section, we will give a brief description of the $f(T)$
gravity and its field equations. The action of $f(T)$ gravity as a
generalization of teleparallel gravity is given by the following relation

\begin{equation}
S=\int \left[ T+f(T)+L_{matter}\right] ed^{4}x,  \label{eqn1}
\end{equation}%
where $T$ is the torsion scalar, $f(T)$ is a general differentiable function
of torsion, $L_{matter}$ is the matter Lagrangian density and $e=\sqrt{-g}%
=\det \left[ e_{\mu }^{i}\right] $. The torsion scalar is defined as

\begin{equation}
T=T_{\mu \nu }^{\alpha }S_{\alpha }^{\mu \nu },  \label{eqn2}
\end{equation}%
where $S_{\alpha }^{\mu \nu }$ and torsion tensor $T_{\mu \nu }^{\alpha }$
are given as follows

\begin{equation}
S_{\alpha }^{\mu \nu }=\frac{1}{2}\left( K_{\alpha }^{\mu \nu }+\delta
_{\alpha }^{\mu }T_{\beta }^{\beta \nu }-\delta _{\alpha }^{\nu }T_{\beta
}^{\beta \mu }\right) ,  \label{eqn3}
\end{equation}

\begin{equation}
T_{\mu \nu }^{\alpha }=\Gamma _{\mu \nu }^{\alpha }-\Gamma _{\nu \mu
}^{\alpha }=e_{i}^{\alpha }\left( \partial _{\mu }e_{\nu }^{i}-\partial
_{\nu }e_{\mu }^{i}\right) .  \label{eqn4}
\end{equation}

In the previous equation, $e_{\mu }^{i}$ represents the components of the
non-trivial tetrad field $e_{i}$ in the coordinate base. One chooses at
random the tetrad field related to the metric tensor $g_{\mu \nu }$ by the
following relation

\begin{equation}
g_{\mu \nu }=\eta _{ij}e_{\mu }^{i}e_{\nu }^{j},  \label{eqn5}
\end{equation}%
where $\eta _{ij}=diag\left( 1,-1,-1,-1\right) $ is the Minkowski space-time
metric and $e_{\mu }^{i}e_{j}^{\mu }=\delta _{j}^{i}$ or $e_{\mu
}^{i}e_{i}^{\nu }=\delta _{\mu }^{\nu }$. The process for evaluating the
tetrad field has been provided in \cite{ref35, ref36, ref37}%
. Note that the Latin alphabets $(i,j,...=1,2,3)$ will be used to denote the
indices of the tetrad field and the Greek alphabets $(\mu ,\nu ,...=0,1,2,3)$
to denote the space-time indices.

The contortion tensor $K_{\alpha }^{\mu \nu }$\ is defined as the difference
between the Levi-Civita and Weitzenb\"{o}ck connections, it is given as
follows

\begin{equation}
K_{\alpha }^{\mu \nu }=-\frac{1}{2}\left( T_{\alpha }^{\mu \nu }-T_{\alpha
}^{\nu \mu }-T_{\alpha }^{\mu \nu }\right) .  \label{eqn6}
\end{equation}

The functional variation of the action in Eq. (\ref{eqn1}) with respect to
tetrads leads to the following field equations

\begin{equation}
S_{\mu }^{\nu \rho }\partial _{\rho }Tf_{TT}+\left[ e^{-1}e_{\mu
}^{i}\partial _{\rho }\left( ee_{i}^{\alpha }S_{\alpha }^{\nu \rho }\right)
+T_{\lambda \mu }^{\alpha }S_{\alpha }^{\nu \lambda }\right] \left(
1+f_{T}\right) +\frac{1}{4}\delta _{\mu }^{\nu }\left( T+f\right)
=k^{2}T_{\mu }^{\nu }.  \label{eqn7}
\end{equation}

Here, $f_{T}=\frac{df\left( T\right) }{dT}$, $f_{TT}=\frac{d^{2}f\left(
T\right) }{dT^{2}}$, $k^{2}=8\pi G=1$ and $T_{\mu }^{\nu }$ is the
energy-momentum tensor of matter. The field equation (\ref{eqn7}) for $f(T)$
gravity is written in terms of tetrad and partial derivatives and appears
very different from Einstein's equations in GR. If we consider the function $%
f(T)=constant$ this leads to Einstein's ordinary field equations.

\section{Metric and field equations}

There are eleven types of Bianchi metrics (I-IX). In this article, we will
study the Bianchi type-I (B-I) metric, which is one of the simplest models
of the spatially homogeneous and anisotropic Universe. This metric is a
direct generalization of the spatially homogeneous and isotropic FRW metric
with a scale factor in each direction and is given as follows

\begin{equation}
ds^{2}=dt^{2}-A^{2}\left( t\right) dx^{2}-B\left( t\right) ^{2}\left[
dy^{2}+dz^{2}\right] ,  \label{eqn8}
\end{equation}%
where $A$ and $B$ are functions of cosmic time $t$ only. The corresponding
torsion scalar is given by

\begin{equation}
T=-2\left( 2\frac{\overset{.}{A}}{A}\frac{\overset{.}{B}}{B}+\frac{\overset{.%
}{B}^{2}}{B^{2}}\right) .  \label{eqn9}
\end{equation}

The energy-momentum tensor $T_{\mu }^{\nu }$ for the perfect fluid
distribution can be represented as

\begin{equation}
T_{\mu }^{\nu }=\left( p+\rho \right) u^{\nu }u_{\mu }-pg_{\mu }^{\nu },
\label{eqn10}
\end{equation}%
where $p$ and $\rho $ are the pressure and energy density of the cosmic
fluid respectively and $u^{\nu }=\left( 0,0,0,1\right) $ is the
four-velocity vector satisfying $u^{\nu }u_{\nu }=1$.

Now, using the field equation (\ref{eqn7}) for the Bianchi type-I metric (%
\ref{eqn8}) and perfect fluid distribution in Eq. (\ref{eqn10}), the
modified Friedmann field equations are given by

\begin{equation}
\left( T+f\right) +4\left( 1+f_{T}\right) \left\{ \frac{\overset{..}{B}}{B}+%
\frac{\overset{.}{B}^{2}}{B^{2}}+\frac{\overset{.}{A}}{A}\frac{\overset{.}{B}%
}{B}\right\} +4\frac{\overset{.}{B}}{B}\overset{.}{T}f_{TT}=-p\left(
t\right) ,  \label{eqn11}
\end{equation}

\begin{equation}
\left( T+f\right) +2\left( 1+f_{T}\right) \left\{ \frac{\overset{..}{A}}{A}+%
\frac{\overset{..}{B}}{B}+\frac{\overset{.}{B}^{2}}{B^{2}}+3\frac{\overset{.}%
{A}}{A}\frac{\overset{.}{B}}{B}\right\} +2\left\{ \frac{\overset{.}{A}}{A}+%
\frac{\overset{.}{B}}{B}\right\} \overset{.}{T}f_{TT}=-p\left( t\right) ,
\label{eqn12}
\end{equation}

\begin{equation}
\left( T+f\right) +4\left( 1+f_{T}\right) \left\{ \frac{\overset{.}{B}^{2}}{%
B^{2}}+2\frac{\overset{.}{A}}{A}\frac{\overset{.}{B}}{B}\right\} =\rho
\left( t\right) .  \label{eqn13}
\end{equation}%
where the dot $\left( .\right) $ denotes the derivative with respect to time 
$t$.

The field equations (\ref{eqn11})-(\ref{eqn13}) are a set of three
differential equations that contain five unknowns $A$, $B$, $f\left(
T\right) $, $p\left( t\right) $, $\rho \left( t\right) $. In order to solve
the field equations explicitly, we need two additional constraints which we
will assume in the next section. Now we will know some of the physical and
geometric quantities that we will need later.

The mean scale factor $a$ of the Bianchi type-I Universe is given by

\begin{equation}
a=\left( AB^{2}\right) ^{\frac{1}{3}}.  \label{eqn14}
\end{equation}

The spatial volume $V$\ of the Universe is defined as%
\begin{equation}
V=a^{3}=AB^{2}.  \label{eqn15}
\end{equation}

Now, the directional Hubble parameters $H_{i}\left( i=1,2,3\right) $ are
respectively

\begin{equation}
H_{1}=\frac{\overset{.}{A}}{A},\text{ \ \ \ \ }H_{2}=\frac{\overset{.}{B}}{B}%
\text{ \ \ and \ \ }H_{2}=H_{3}.  \label{eqn16}
\end{equation}

The mean Hubble's parameter is defined as

\begin{equation}
H=\frac{1}{3}\left( H_{1}+2H_{2}\right) .  \label{eqn17}
\end{equation}

Using Eqs. (\ref{eqn14})-(\ref{eqn17}), we find

\begin{equation}
H=\frac{1}{3}\frac{\overset{.}{V}}{V}=\frac{1}{3}\left( \frac{\overset{.}{A}%
}{A}+2\frac{\overset{.}{B}}{B}\right) .  \label{eqn18}
\end{equation}

Other physical parameters, the expansion scalar $\theta $, the mean
anisotropic parameter $A_{m}$ and the shear scalar $\sigma ^{2}$, are
defined for the Bianchi type-I Universe, as

\begin{equation}
\theta =\frac{\overset{.}{A}}{A}+2\frac{\overset{.}{B}}{B}=3H,  \label{eqn19}
\end{equation}

\begin{equation}
A_{m}=\frac{2}{3}\frac{\sigma^{2}}{H^{2}} =\frac{1}{3}{\sum} \left( \frac{\Delta H_{i}}{H}\right) ^{2},  \label{eqn20}
\end{equation}

\begin{equation}
\sigma ^{2}=\frac{1}{2}\left[ \left( \frac{\overset{.}{A}}{A}\right)
^{2}+2\left( \frac{\overset{.}{B}}{B}\right) ^{2}\right] -\frac{\theta ^{2}}{%
6},  \label{eqn21}
\end{equation}%
where $\Delta H_{i}=H_{i}-H$ and $H_{i}\left( i=1,2,3\right) $ represent the
directional Hubble parameters.

\section{Solutions of the field equations}

In this section, we find exact solutions of field equations using the linear
form of $f(T)$ gravity, i.e.

\begin{equation}
f(T)=T.  \label{eqn22}
\end{equation}

Using Eqs. (\ref{eqn11}) and (\ref{eqn12}) yields

\begin{equation}
\frac{d}{dt}\left( \frac{\overset{.}{A}}{A}-\frac{\overset{.}{B}}{B}\right)
+\left( \frac{\overset{.}{A}}{A}-\frac{\overset{.}{B}}{B}\right) \frac{%
\overset{.}{V}}{V}=0.  \label{eqn23}
\end{equation}

By integrating the previous equation, we get

\begin{equation}
\frac{A}{B}=k_{2}\exp \left[ k_{1}\int \frac{dt}{V}\right] ,  \label{eqn24}
\end{equation}%
where $k_{1}$ and $k_{2}$ are constants of integration.

Using Eq. (\ref{eqn15}), we get the metric potentials as follows

\begin{equation}
A=D_{1}V^{\frac{1}{3}}\exp \left( \chi _{1}\int \frac{1}{V}dt\right) ,
\label{eqn25}
\end{equation}

\begin{equation}
B=D_{2}V^{\frac{1}{3}}\exp \left( \chi _{2}\int \frac{1}{V}dt\right) ,
\label{eqn26}
\end{equation}%
where $D_{i}\left( i=1,2\right) $ and $\chi _{i}\left( i=1,2\right) $
satisfy the relation $D_{1}D_{2}^{2}=1$ and $\chi _{1}+2\chi _{2}=0$. To
complete the additional constraints to find the exact solutions to the field
equations, we reduce the second constraint on the scale factor or in some
other way the deceleration parameter (DP).

The DP is given by

\begin{equation}
q=-\frac{a\overset{..}{a}}{\overset{.}{a}^{2}}=\frac{d}{dt}\left( \frac{1}{H}%
\right) -1.  \label{eqn27}
\end{equation}

The DP is an important tool to describe the evolution of the Universe. If $%
q>0$ indicates the cosmic deceleration while $q<0$ shows the cosmic
acceleration. According to recent observational data of the SNIa, our
Universe goes from the initial deceleration phase to the current
acceleration phase, that is, it goes from a positive value of the DP $\left(
q>0\right) $ to a negative value $\left( q<0\right) $, indicating that the
DP is a function that varies with cosmic time $t$. Therefore, in this
article, we assume that the DP varies with cosmic time as a linear function
of the Hubble parameter $H$ \cite{ref38}

\begin{equation}
q=\alpha +\beta H.  \label{eqn28}
\end{equation}

Here $\alpha $\ and $\beta $\ are arbitrary constants. We solve Eq. (\ref%
{eqn28}) for $\alpha =-1$, we find the scale factor as follows

\begin{equation}
a\left( t\right) =\exp \left( \frac{1}{\beta }\sqrt{2\beta t+c}\right) ,
\label{eqn29}
\end{equation}%
where $c$ is an integrating constant, and for the rest of the article, we'll
use that $F=\frac{1}{\beta }\sqrt{2\beta t+c}$.

Similarly, the Hubble parameter $H$ and DP $q$ in terms of cosmic time $t$
are obtained as

\begin{equation}
H=\frac{1}{\beta F},  \label{eqn30}
\end{equation}

\begin{equation}
q=-1+\frac{1}{F}.  \label{eqn31}
\end{equation}

The choice $\alpha =-1$ is appropriate to obtain a Hubble parameter which
depends on cosmic time instead of being constant \cite{ref39, ref40, ref41,
ref42, ref43}. Also, we use $\alpha =-1$ to get the time-dependent DP, but
if we choose $\alpha \neq -1$, we will find that the DP takes a constant
value $q=-1$. By Eqs. (\ref{eqn30}) and (\ref{eqn31}), we show that $%
H\rightarrow 0$ and $q\rightarrow -1$ as $t\rightarrow \infty $. Moreover, $%
q\geq 0$ for $t\leq \frac{\beta }{2}-\frac{c}{2\beta }$ and $q<0$ for $t>%
\frac{\beta }{2}-\frac{c}{2\beta }$. To study the behavior of certain
cosmological parameters in terms of redshift $z$, we must first give the
relation between the redshift $z$\ and the scale factor $a\left( t\right) $,
which is written as follows

\begin{equation}
z=\frac{a\left( t_{0}\right) }{a\left( t\right) }-1,  \label{eqn32}
\end{equation}%
where $a\left( t_{0}\right) $ is the current value of scale factor.

Using Eqs. (\ref{eqn29}) and (\ref{eqn32}), we find the cosmic time $t$ in
terms of redshift $z$ as

\begin{equation}
t\left( z\right) =\frac{\beta }{2}\left[ \left\{ F_{0}-\log \left(
1+z\right) \right\} ^{2}-\frac{c}{\beta ^{2}}\right] ,  \label{eqn33}
\end{equation}%
where, $F_{0}=\frac{1}{\beta }\sqrt{2\beta t_{0}+c}$ and $t_{0}$ denotes the
present time.

With a simple calculation, we find the Hubble parameter $H$ and the DP $q$
in terms of redshift $z$\ as follows

\begin{equation}
H\left( z\right) =\frac{1}{\beta \left( F_{0}-\log \left( 1+z\right) \right) 
},  \label{eqn34}
\end{equation}

\begin{equation}
q\left( z\right) =-1+\frac{1}{F_{0}-\log \left( 1+z\right) }.  \label{eqn35}
\end{equation}

From Eqs. (\ref{eqn34}) and (\ref{eqn35}), it is clear that $H\rightarrow 0$
and $q\rightarrow -1$ as $z\rightarrow -1$. In Tab. 1, we summarize the
dynamics of the Universe for $a\left( t\right) =\exp \left( F\right) $. Now,
in Fig. 1, we plot the behavior of the Hubble parameter $H$ and the
deceleration parameter $q$ in terms of redshift $z$ for the value of the
pair $\left( \beta ,c\right) $ as $\left( 3,2.85\right) $, respectively. Our
model is transforming from $q>0$ (deceleration) to $q<0$ (acceleration)
phases. According to observational data, the DP $q$ value lies between the
range $-1<q<0$ and the expansion of the current Universe is accelerating.
Therefore, the current value of the deceleration parameter is consistent
with recent observations i.e. $q_{0}=-0.68$. In the following, we have given
the values of the deceleration parameter and the Hubble parameter for the
current time according to some observational data:

\begin{itemize}
\item \textbf{Case I Based on SNIa union data }\cite{ref44}:\textbf{\ }Based
on these observational data, $q_{0}=-0.73$ and $H_{0}=73.8$.

\item \textbf{Case II Based on SNIa data in combination with BAO and CMB
observations }\cite{ref45}:\textbf{\ }In that case, $q_{0}=-0.54$ and $%
H_{0}=73.8$.

\item \textbf{Case III Based on current data in combination with OHD and JLA
observations }\cite{ref46}:\textbf{\ }In that case, $q_{0}=-0.52$ and $%
H_{0}=69.2$.
\end{itemize}

\begin{table}[tbp]
\begin{center}
\begin{tabular}{ccccc}
\hline\hline
Time $\left( t\right) $ & Redshift $\left( z\right) $ & $a$ & $q$ & $H$ \\ 
\hline\hline
$t\rightarrow 0$ & $z\rightarrow \exp \left[ \frac{1}{\beta }\left( \sqrt{%
2\beta t_{0}+c}-\sqrt{c}\right) \right] $ & $\exp \left( \frac{\sqrt{c}}{%
\beta }\right) $ & $-1+\frac{\beta }{\sqrt{c}}$ & $\frac{1}{\sqrt{c}}$ \\ 
\hline\hline
$t\rightarrow \infty $ & $z\rightarrow -1$ & $\infty $ & $-1$ & $0$ \\ 
\hline\hline
\end{tabular}%
\end{center}
\caption{Dynamics of the Universe for $a\left( t\right) =\exp \left( \frac{1%
}{\protect\beta }\protect\sqrt{2\protect\beta t+c}\right) $.}
\label{tab1}
\end{table}

\begin{figure*}
\centerline{\includegraphics[scale=1]{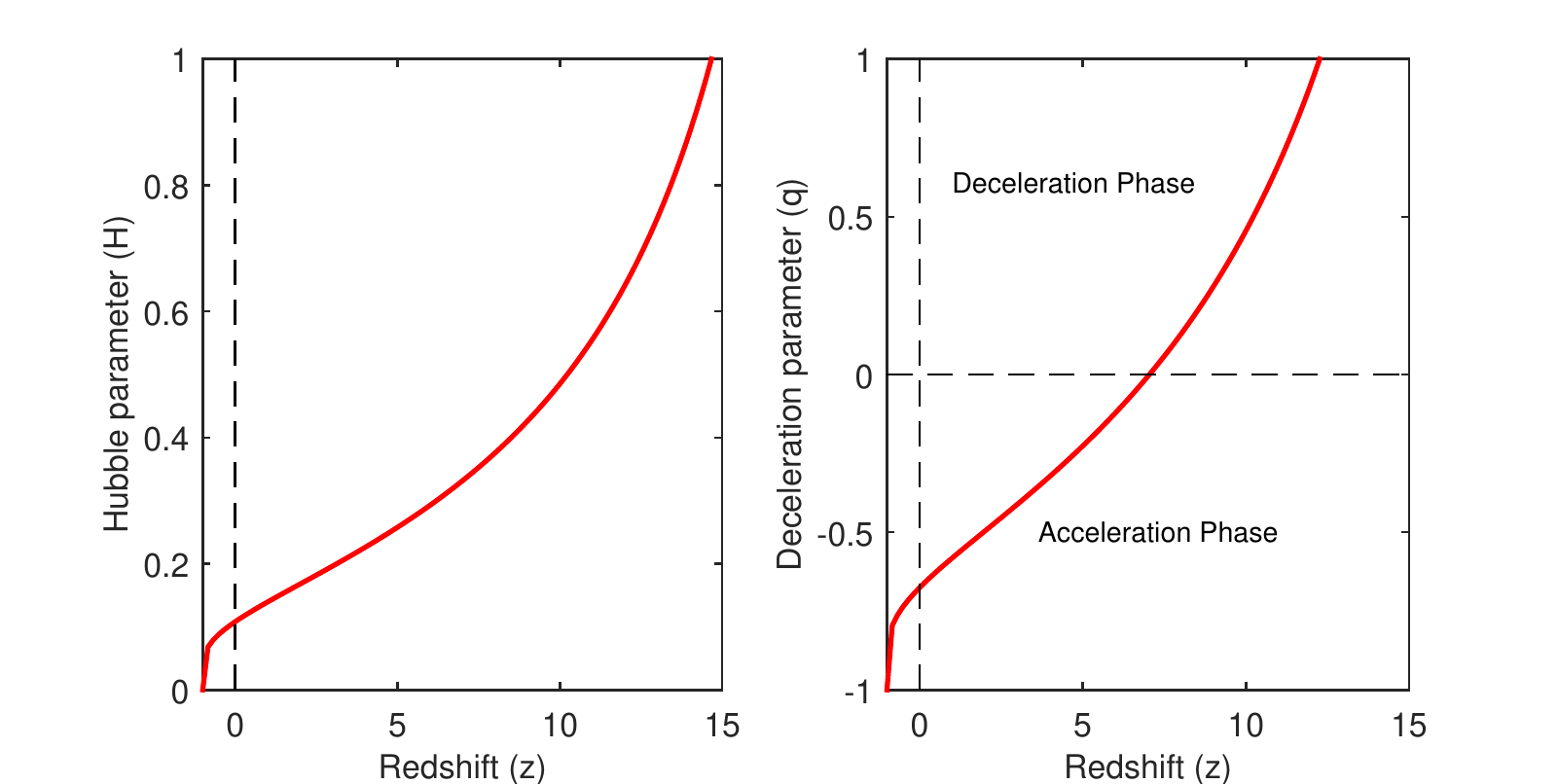}}
\caption{{\emph{(left) The plot of $H$ versus $z$, (right) The plot of $q$
versus $z$.}}}
\label{fig1}
\end{figure*}

Using Eqs. (\ref{eqn29}) in (\ref{eqn25}) and (\ref{eqn26}), we obtain the
metric potentials as follows

\begin{equation}
A\left( t\right) =D_{1}\exp \left( F\right) \exp \left[ -\frac{\chi _{1}}{9}%
\beta \left( 3F+1\right) \exp \left( -3F\right) \right] ,  \label{eqn36}
\end{equation}%
and

\begin{equation}
B\left( t\right) =D_{2}\exp \left( F\right) \exp \left[ -\frac{\chi _{2}}{9}%
\beta \left( 3F+1\right) \exp \left( -3F\right) \right] .  \label{eqn37}
\end{equation}

Using Eqs. (\ref{eqn32}) and (\ref{eqn33}), the metric in Eq. (\ref{eqn8})
becomes as follows

\begin{eqnarray}
ds^{2} &=&dt^{2}-D_{1}^{2}\exp \left( 2F\right) \exp \left[ -\frac{2\chi _{1}%
}{9}\beta \left( 3F+1\right) \exp \left( -3F\right) \right] dx^{2}
\label{eqn38} \\
&&-D_{2}^{2}\exp \left( 2F\right) \exp \left[ -\frac{2\chi _{2}}{9}\beta
\left( 3F+1\right) \exp \left( -3F\right) \right] \left[ dy^{2}+dz^{2}\right]
.  \notag
\end{eqnarray}%
\newline

The torsion scalar $T$\ for the model becomes

\begin{equation}
T=\frac{-6}{\beta ^{2}F^{2}}-2\chi _{2}\left( 2\chi _{1}+\chi _{2}\right)
\exp \left( -6F\right) .  \label{eqn39}
\end{equation}

\section{Physical and geometrical properties of the model}

The directional\ Hubble parameters, which determine the expansion rate of
the Universe, are given by

\begin{equation}
H_{1}=\chi _{1}\exp \left( -3F\right) +\frac{1}{\beta F},  \label{eqn40}
\end{equation}

\begin{equation}
H_{2}=H_{3}=\chi _{2}\exp \left( -3F\right) +\frac{1}{\beta F}.
\label{eqn41}
\end{equation}

The expansion scalar $\theta $ and the shear $\sigma ^{2}$ are obtained as

\begin{equation}
\theta =3H=\frac{3}{\beta F},  \label{eqn42}
\end{equation}

\begin{equation}
\sigma ^{2}=\frac{1}{2}\left( \frac{\chi _{1}^{2}+2\chi _{2}^{2}}{a^{6}}%
\right) =\frac{1}{2}\left( \chi _{1}^{2}+2\chi _{2}^{2}\right) \exp \left(
-6F\right) .  \label{eqn43}
\end{equation}

Using Eq. (\ref{eqn29}) into Eq. (\ref{eqn15}) we get the spatial volume as

\begin{equation}
V=\exp \left( 3F\right) .  \label{eqn44}
\end{equation}

The average anisotropy parameter $A_{m}$ is given as

\begin{equation}
A_{m}=\frac{1}{3}\left( \chi _{1}^{2}+2\chi _{2}^{2}\right) \beta
^{2}F^{2}\exp \left( -6F\right) .  \label{eqn45}
\end{equation}

From Eq. (\ref{eqn44}), it is clear that the spatial volume of the model is
finite at the initial singularity (i.e. at $t=0$) and approaches infinity as 
$t\rightarrow \infty $. Moreover, the average scale factor $a\left( t\right) 
$ in Eq. (\ref{eqn29}) is also finite at the early epoch of the Universe. It
shows that the obtained model of the Universe is expanding continuously with
cosmic time $t$. Eqs. (\ref{eqn40})-(\ref{eqn43}) show the directional\
Hubble parameters $H_{i}$, the scalar expansion $\theta $ and the scalar
shear $\sigma ^{2}\rightarrow 0$ as $t\rightarrow \infty $ and they approach
finite value as $t\rightarrow 0$. Finally, from Eq. (\ref{eqn45}), we
observe that the average anisotropy parameter $A_{m}\rightarrow 0$ as $%
t\rightarrow \infty $. This indicates that our model contains a transition
from the early anisotropic Universe to the current isotropic Universe as
shown by observational data.

Using Eqs. (\ref{eqn36}) and (\ref{eqn37}) in the field equations (\ref%
{eqn11})-(\ref{eqn13}), with simple math, the physical parameters such as
energy density $\rho \left( t\right) $, cosmic pressure $p\left( t\right) $
are obtained as

\begin{equation}
\rho \left( t\right) =\frac{12}{\beta ^{2}F^{2}}+4\chi _{2}\left( 2\chi
_{1}+\chi _{2}\right) \exp \left( -6F\right)  \label{eqn46}
\end{equation}

\begin{equation}
p\left( t\right) =\frac{8}{\beta ^{2}F^{3}}+\frac{24\chi _{2}}{\beta F}\exp
\left( -3F\right) -12\left\{ \chi _{2}\exp \left( -3F\right) +\frac{1}{\beta
F}\right\} ^{2}  \label{eq47}
\end{equation}

Using the relationship between cosmic time $t\left( z\right) $\ and redshift 
$z$ in Eq. (\ref{eqn33}) and Eqs. (\ref{eqn46}) and (\ref{eq47}), we plot
the behavior of the energy density $\rho \left( z\right) $\ and pressure $%
p\left( z\right) $\ of the Universe versus redshift $z$ in Fig. 2,
respectively. First of all, note that $\rho \left( z\right) $ and $p\left(
z\right) $ $\rightarrow 0$ as $z\rightarrow -1$ (or $t\rightarrow \infty $),
which is similar behavior to the big-bang model. From Fig. 2 (left), we can
observe that the energy density remains positive throughout the evolution of
the Universe and is a decreasing function of redshift $z$. The pressure in
Fig. 2 (right), evolves from early positive values to present negative ones.
As per the observation, the negative pressure is due to DE in the context of
accelerated expansion of the Universe. Hence, the behavior of pressure in
our model is consistent with this observation.

\begin{figure*}
\centerline{\includegraphics[scale=1]{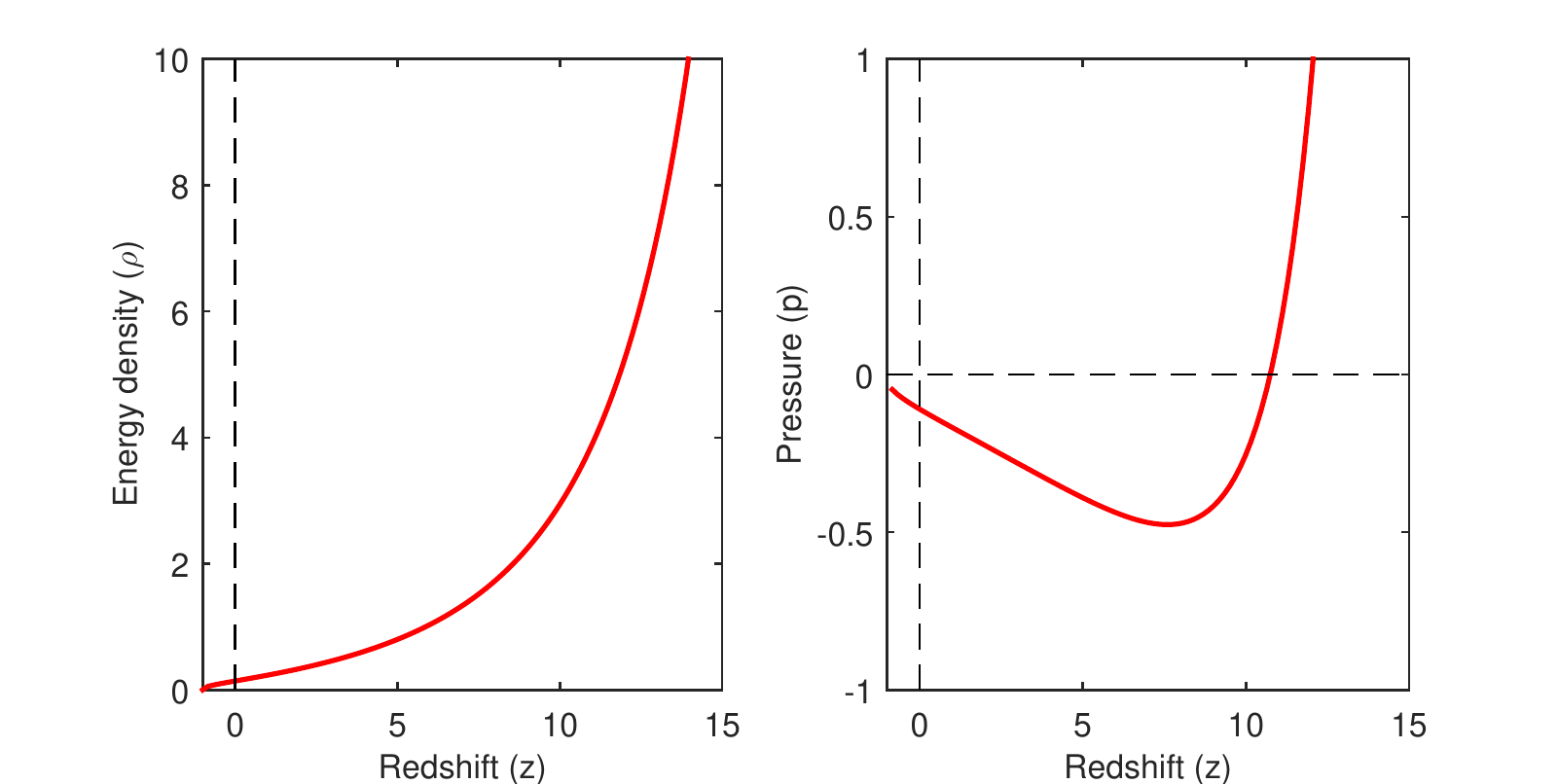}}
\caption{{\emph{(left) The plot of $\protect\rho $ versus $z$, (right) The
plot of $p$ versus $z$.}}}
\label{fig2}
\end{figure*}

Using the equation of state for a perfect fluid $\left( \omega =\frac{p}{%
\rho }\right) $, and using Eqs. (\ref{eqn46}) and (\ref{eq47}), we find the
EoS parameter as follows

\begin{equation}
\omega \left( t\right) =\frac{\frac{2}{\beta ^{2}F^{3}}+\frac{6\chi _{2}}{%
\beta F}\exp \left( -3F\right) -3\left\{ \chi _{2}\exp \left( -3F\right) +%
\frac{1}{\beta F}\right\} ^{2}}{\frac{3}{\beta ^{2}F^{2}}+\chi _{2}\left(
2\chi _{1}+\chi _{2}\right) \exp \left( -6F\right) }  \label{eqn48}
\end{equation}

The EoS parameter is among the basic tools for studying the different phases
of the Universe as well as the history of the Universe. If $\omega =-1$, it
represents $\Lambda CDM$ model, $-1<\omega <-\frac{1}{3}$, represents
quintessence model and $\omega <-1$, indicates phantom behavior of the
model. We have plotted the EoS parameter $\omega \left( z\right) $ for
redshift $z$ in Fig. 3 (left) for a fixed value of the pair $\left( \beta
,c\right) $, the EoS parameter $\omega \in $ quintessence region for high
redshift $z$ and over time, $\omega \rightarrow -1$ in infinite future (i.e. 
$z\rightarrow -1$). The present value of the EoS parameter of our model is
consistent with the observational data on $\omega $ from Planck data \cite%
{ref47}:

\begin{itemize}
\item $\omega =-1.56_{-0.84}^{+0.60}$ (Planck + TT + lowE),

\item $\omega =-1.58_{-0.41}^{+0.52}$ (Planck + TT, EE + lowE),

\item $\omega =-1.57_{-0.40}^{+0.50}$ (Planck + TT, TE, EE + lowE + lensing),

\item $\omega =-1.04_{-0.10}^{+0.10}$ (Planck + TT, TE, EE + lowE + lensing
+ BAO).
\end{itemize}

From Fig. 3 (left), it is clear that the EoS parameter of our model is
within the range of the above observational data. Accordingly, our results
are in agreement with previous observational data.

\begin{figure*}
\centerline{\includegraphics[scale=1]{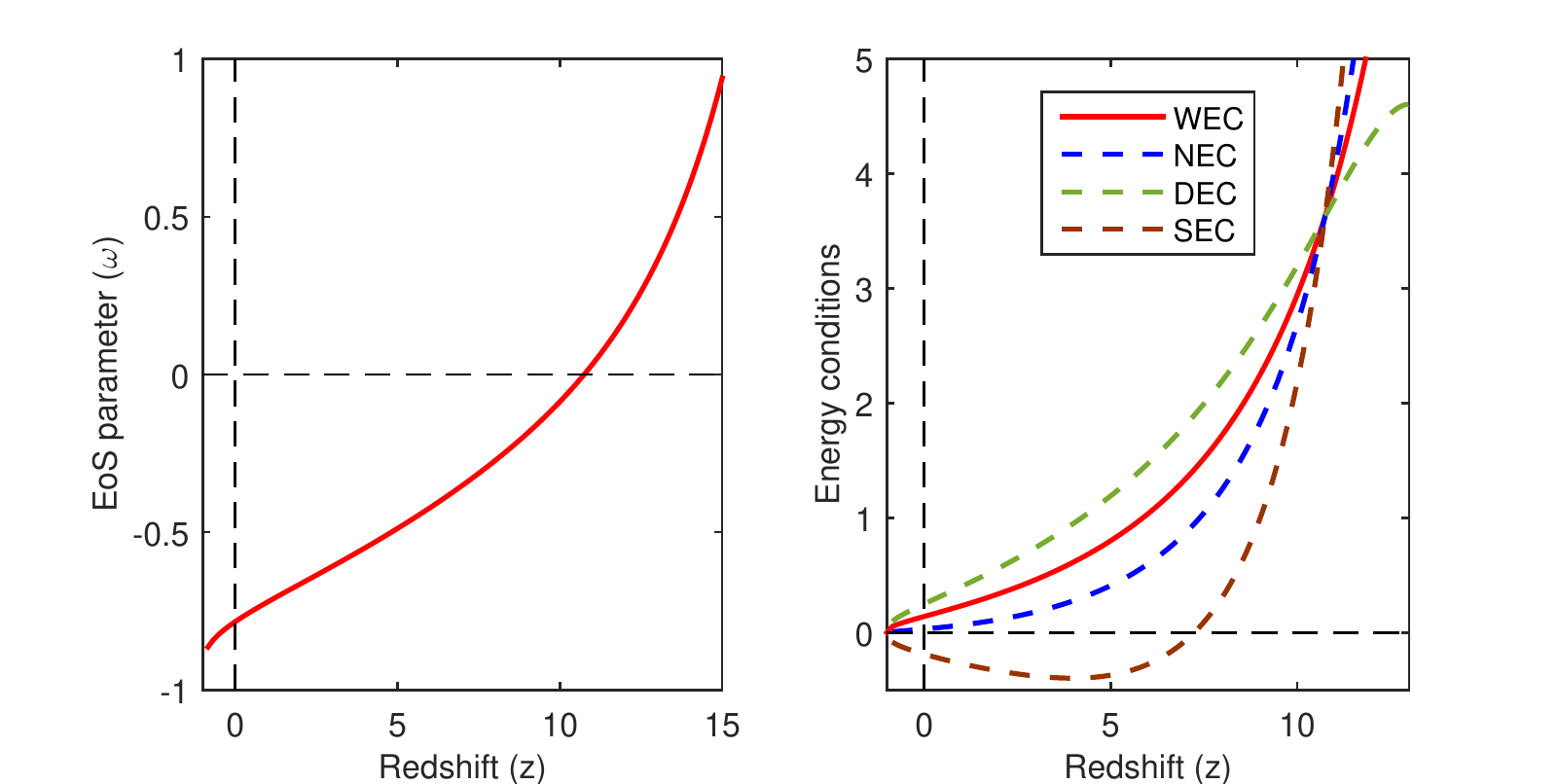}}
\caption{{\emph{(left) The plot of $\protect\omega $ versus $z$, (right) The
plot of energy conditions versus $z$.}}}
\label{fig3}
\end{figure*}

\subsection{Energy Conditions}

Energy conditions are a set of conditions that describe matter in the
Universe and are used in many approaches to understanding the evolution of
the Universe. The role of energy conditions is to verify the acceleration of
the expansion of the Universe. There are many forms of energy conditions
such as null energy condition (NEC), weak energy condition (WEC), dominant
energy condition (DEC), and strong energy condition (SEC). Here \cite{ref48,
ref49, ref50, ref51, ref52}, a group of authors who have done work on energy
conditions. In $f(T)$\ gravity with known energy density and pressure, these
energy conditions are given as follows

\begin{itemize}
\item WEC: $\rho \geq 0$

\item NEC: $\rho +p\geq 0$

\item DEC: $\rho -p\geq 0$

\item SEC: $\rho +3p\geq 0$
\end{itemize}

Fig. 3 (right) represents the energy conditions as a function of time for
our model under study, i.e. the Bianchi type-I Universe with the DP varies
with cosmic time as a linear function of the Hubble parameter $H$. From the
figure below, we notice that the WEC, NEC, and DEC are well satisfied
throughout the cosmic evolution, while there is a clear violation of the
SEC. Thus, the violation of the SEC gives us the acceleration of the
Universe.

\subsection{Perturbation and stability of the obtained solution}

To study the stability of our solutions, we will follow the same approach
found in this work \cite{ref53, ref54}. We will use the perturbation
approach to check the obtained expanding background solution stability
against perturbation of scale factors or the metric field. Now, we will
consider the existence of a perturbation for the three scale factors $%
a_{i}\left( i=1,2,3\right) $ as

\begin{equation}
a_{i}\longrightarrow a_{B_{i}}+\delta a_{i}=a_{B_{i}}\left( 1+\frac{\delta
a_{i}}{a_{B_{i}}}\right) =a_{B_{i}}\left( 1+\delta b_{i}\right) ,
\label{eqn49}
\end{equation}%
where $\delta b_{i}=\frac{\delta a_{i}}{a_{B_{i}}}$.

In the same way, we write the perturbation in the spatial volume $%
V=\prod\nolimits_{i=1}^{3}a_{i}$, directional Hubble parameters $H_{i}=%
\frac{\overset{.}{a_{i}}}{a_{i}}$ and mean Hubble parameter $H=\frac{1}{3}%
\sum_{i=1}^{3}H_{i}$ as follows

\begin{equation}
V\longrightarrow V_{B}+V_{B}\underset{i}{\sum }\delta b_{_{i}},
\label{eqn50}
\end{equation}

\begin{equation}
H_{i}\longrightarrow H_{B_{i}}\underset{i}{+\sum }\delta b_{_{i}},
\label{eqn51}
\end{equation}

\begin{equation}
H\longrightarrow H_{B}+\frac{1}{3}\underset{i}{\sum }\delta b_{_{i}}.
\label{eqn52}
\end{equation}

Here, $V_{B}$, $H_{B_{i}}$ and $H_{B}$ are the background spatial volume,
directional Hubble parameters, and mean Hubble parameter respectively. Now,
it can be shown that the metric linear order perturbations $\delta b_{_{i}}$%
\ satisfy the following differential equations

\begin{equation}
\underset{i}{\sum }\delta \overset{..}{b}_{i}+2\underset{i}{\sum }%
H_{B_{i}}\delta \overset{.}{b}_{i}=0,  \label{eqn53}
\end{equation}

\begin{equation}
\delta \overset{..}{b}_{i}+\frac{\overset{.}{V}_{B}}{V_{B}}\delta \overset{.}%
{b}_{i}+\underset{j}{\sum }\delta \overset{.}{b}_{j}H_{B_{i}}=0,
\label{eqn54}
\end{equation}

\begin{equation}
\underset{i}{\sum }\delta \overset{.}{b}_{i}=0.  \label{eqn55}
\end{equation}

With a little math, we can easily find through Eqs. (\ref{eqn53})-(\ref%
{eqn55}) the following relation

\begin{equation}
\delta \overset{..}{b}_{i}+\frac{\overset{.}{V}_{B}}{V_{B}}\delta \overset{.}%
{b}_{i}=0.  \label{eqn56}
\end{equation}

For our model, $V_{B}$ is given by

\begin{equation}
V_{B}=\exp \left( \frac{3}{\beta }\sqrt{2\beta t+c}\right) .  \label{eqn57}
\end{equation}

Using the above condition in Eq. (\ref{eqn56}) and after integration, we find

\begin{equation}
\delta b_{i}=-c_{i}\left[ \frac{\left( \beta +3\sqrt{2\beta t+c}\right) }{%
9\exp \left( \frac{3\sqrt{2\beta t+c}}{\beta }\right) }\right] ,
\label{eqn58}
\end{equation}%
where $c_{i}$ is an integrating constant. Thus, for each scale factor $a_{i}$%
, the actual fluctuations are given by

\begin{equation}
\delta a_{i}=-c_{i}\left[ \frac{\left( \beta +3\sqrt{2\beta t+c}\right) }{%
9\exp \left( \frac{3\sqrt{2\beta t+c}}{\beta }\right) }\right] .
\label{eqn59}
\end{equation}

From the above equation, it is clear that $\delta a_{i}$ approaches zero as $%
t\rightarrow \infty $. The same behavior is illustrated by Fig. 4 (left)
which represents the variation of $\delta a_{i}$ in terms of the redshift $z$%
, i.e. $\delta a_{i}\rightarrow 0$ as $z\rightarrow -1$. Thus, the
background solution is stable against the perturbation of the metric.

\subsection{Jerk parameter}

As it is known in the literature, the jerk parameter is one of the
fundamental physical quantities to describe the dynamics of the Universe.
The Jerk parameter is a dimensionless third derivative of the scale factor $%
a\left( t\right) $ for cosmic time $t$ and is defined as

\begin{equation}
j=\frac{\overset{...}{a}}{aH^{3}}.  \label{eqn60}
\end{equation}

Eq. (\ref{eqn60}) can be written in terms of a DP as

\begin{equation}
j=q+2q^{2}-\frac{\overset{.}{q}}{H}.  \label{eqn61}
\end{equation}

Using Eqs. (\ref{eqn30}) and (\ref{eqn31}), the jerk parameter for our model
is

\begin{equation}
j=\frac{3\beta ^{2}}{2\beta t+c}-\frac{3\beta }{\sqrt{2\beta t+c}}+1.
\label{eqn62}
\end{equation}

To study the behavior of the jerk parameter $j\left( z\right) $, it is
better to express it in terms of redshift $z$

\begin{equation}
j\left( z\right) =\frac{3}{\left( F_{0}-\log \left( 1+z\right) \right) ^{2}}-%
\frac{3}{F_{0}-\log \left( 1+z\right) }+1.  \label{eqn63}
\end{equation}

For the $\Lambda CDM$ model, the value of the jerk parameter is $j=1$. The
Universe shifts from the early deceleration phase to the current
acceleration phase with a positive jerk parameter $j_{0}>0$ and a negative
DP $q_{0}<0$ according to the $\Lambda CDM$ model. Fig. 4 (right) represents
the variation of jerk parameter $j$ versus redshift $z$. It is very clear
from this figure, that the jerk parameter remains positive throughout the
cosmic evolution. The current jerk parameter value $j_{0}\left( z=0\right) $
is positive. As a result, at present $z=0$, our model can be expected to
adopt the behavior of another DE model instead of the $\Lambda CDM$ model $%
j\neq 1$, but in the future $z\rightarrow -1$, our model is similar to the $%
\Lambda CDM$ model $j=1$. For comparison with observation data, the value of
the jerk parameter of our model is within the range of values observed by
the SNIa data $\left( j=1.32_{-1.21}^{+1.37}\right) $ \cite{ref55}, and the
combined results of the SNLS project and the X-ray galaxy cluster distance
measurements $\left( j=0.51_{-2.00}^{+2.55}\right) $ \cite{ref56}.

\begin{figure*}
\centerline{\includegraphics[scale=1]{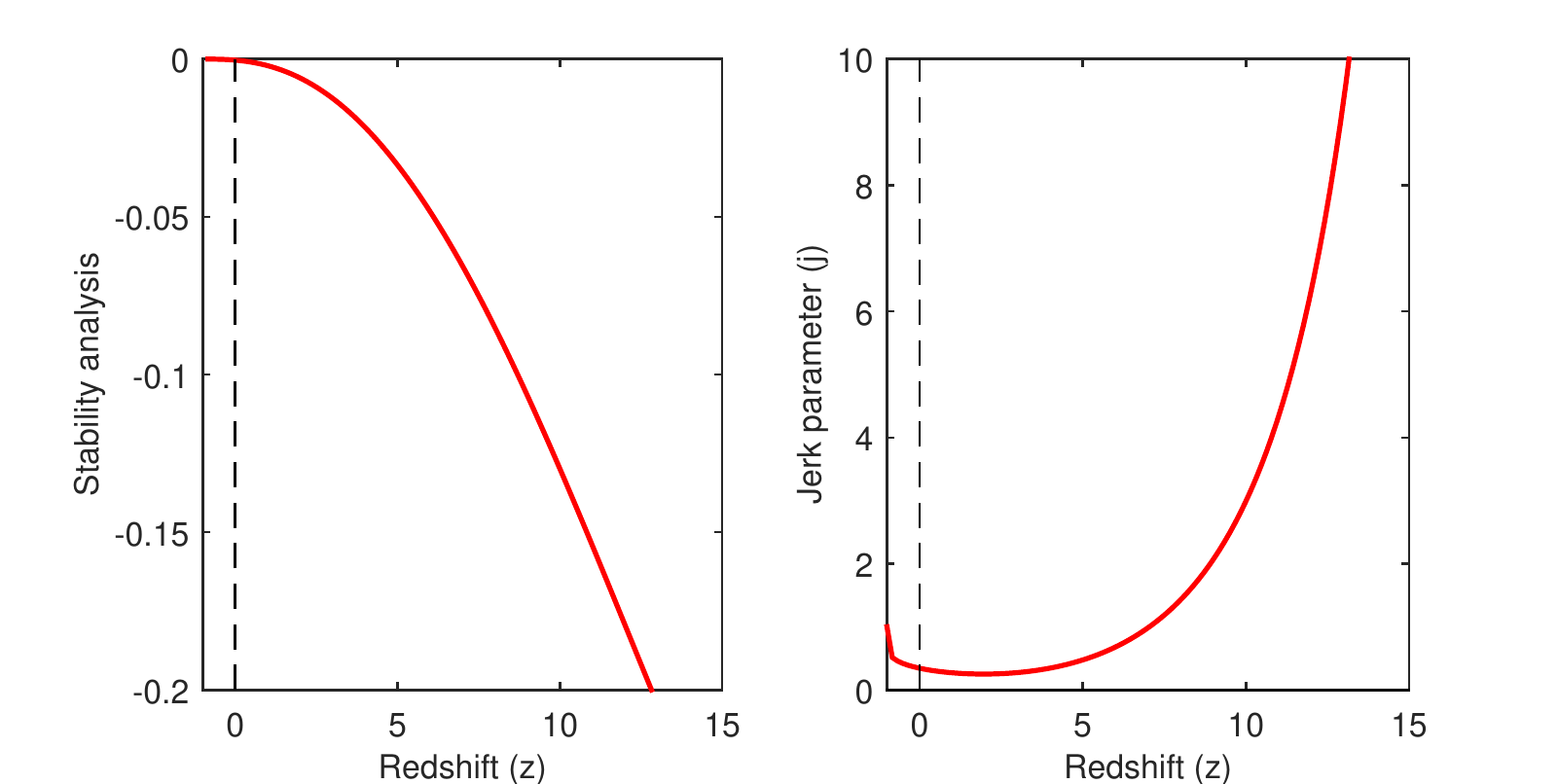}}
\caption{{\emph{(left) The plot of $\protect\delta a_{i}$ versus $z$,
(right) The plot of $j$ versus $z$.}}}
\label{fig4}
\end{figure*}

\subsection{Statefinder diagnostic}

The statefinder pair is a very important geometrical diagnostic tool used to
distinguish different DE models such as $\Lambda CDM$, $HDE$, $CG$, $SCDM$,
and $Quintessence$. The state-finder pair $\left\{ r,s\right\} $\ is defined
as \cite{ref57}

\begin{equation}
r=\frac{\overset{...}{a}}{aH^{3}},\text{ \ \ \ \ }s=\frac{r-1}{3\left( q-%
\frac{1}{2}\right) }  \label{eqn64}
\end{equation}

We can find different models of DE according to the values of the couple $r$
and $s$. In particular,

\begin{itemize}
\item $\Lambda CDM$ corresponds to $\left( r=1,s=0\right) ,$

\item $SCDM$ corresponds to $\left( r=1,s=1\right) ,$

\item $HDE$ corresponds to $\left( r=1,s=\frac{2}{3}\right) ,$

\item $CG$ corresponds to $\left( r>1,s<0\right) ,$

\item $Quintessence$ corresponds to $\left( r<1,s>0\right) $
\end{itemize}

Using Eqs. (\ref{eqn29}), (\ref{eqn30}) and (\ref{eqn31}), the values of the
state-finder parameters in terms of redshift $z$ for our model are

\begin{equation}
r\left( z\right) =\frac{3}{\left( F_{0}-\log \left( 1+z\right) \right) ^{2}}-%
\frac{3}{F_{0}-\log \left( 1+z\right) }+1  \label{eqn65}
\end{equation}

\begin{equation}
s\left( z\right) =\frac{2\left( F_{0}-\log \left( 1+z\right) \right) -2}{%
\left[ 3\left( F_{0}-\log \left( 1+z\right) \right) -2\right] \left(
F_{0}-\log \left( 1+z\right) \right) }  \label{eqn66}
\end{equation}

From Fig. 5, we notice that the statefinder parameters $\left\{ r,s\right\} $
evolve from the CG (Chaplygin Gas) region $\left( r>1,s<0\right) $ to the
quintessence region $\left( r<1,s>0\right) $ at present, and later time to $%
\Lambda CDM$ point $\left( r=1,s=0\right) $. As a result, our model current
behaves like a quintessence model for DE.

\begin{figure}[h]
\centerline{\includegraphics[scale=0.7]{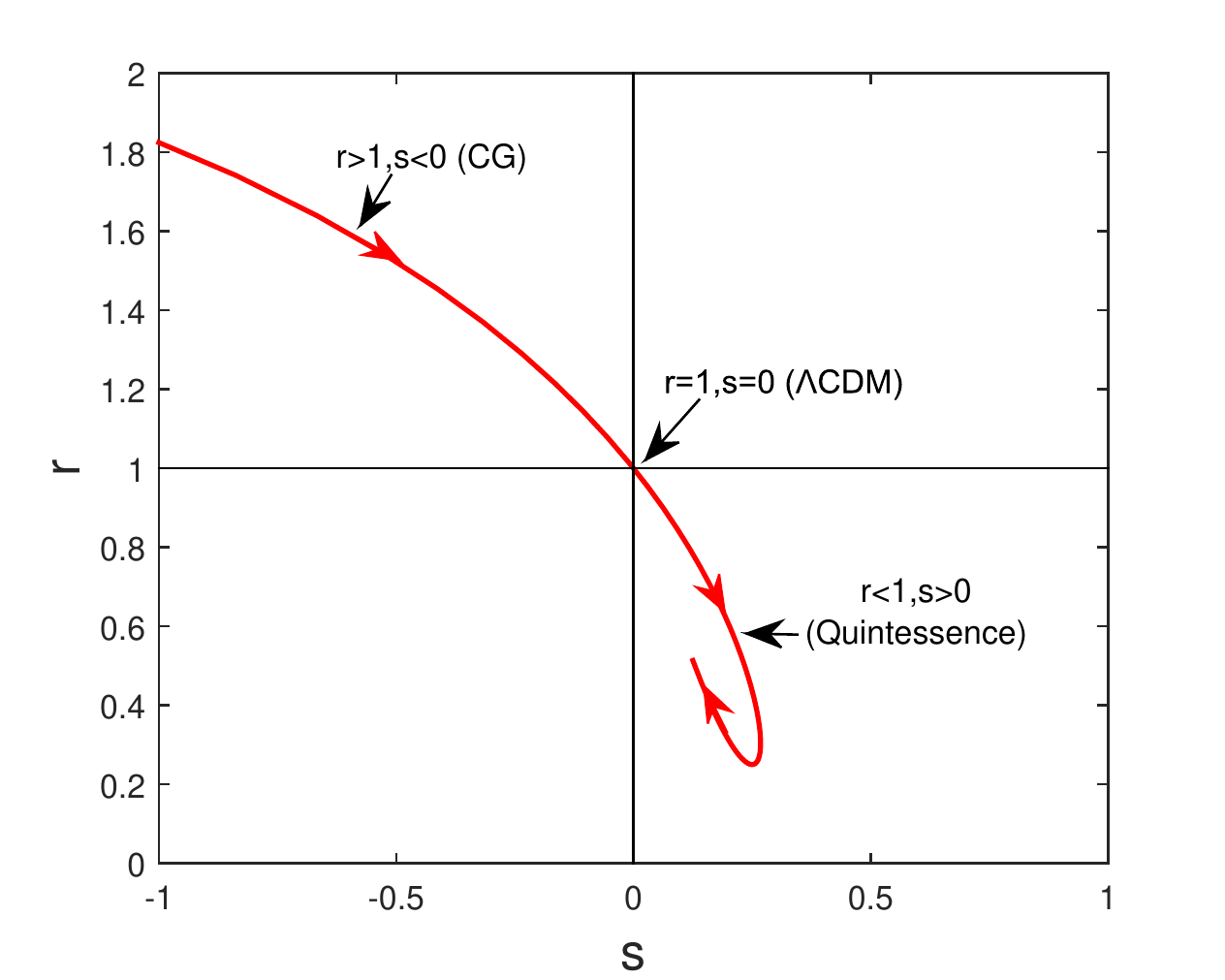}}
\caption{{\emph{The plot of $(s,r)$ trajectories.}}}
\label{fig5}
\end{figure}

\section{Discussions and conclusions}

In this paper, we have studied a cosmological model with a variable
deceleration parameter in a Bianchi type-I Universe in $f(T)$ gravity by
assuming a particular form of the deceleration parameter as a linear
function of the Hubble parameter i.e. $q=-1+\beta H$, $\beta >0$. We
consider $f(T)=T$ and find the field equations for our model and graphically
represent the different physical and geometric parameters as a function of
the redshift. The important results of our model are:

\begin{itemize}
\item The DP of our model gives us two phases of the Universe, the early
deceleration phase, and the current acceleration phase, as indicated by the
observational data. The Hubble parameter of our model is a decreasing
function of redshift $z$. Also, $H\rightarrow 0$ as $z\rightarrow -1$ (i.e. $%
t\rightarrow \infty $).

\item Thus, our model contains a transition from the early anisotropic
Universe $\left( A_{m}\neq 0\right) $ to the current isotropic Universe $%
\left( A_{m}=0\right) $.

\item The energy density of the Universe decreases over time and remains
positive throughout cosmic evolution, while the pressure starts with
positive values then changes to negative values for the current time, the
negative pressure is caused by cosmic acceleration.

\item The EoS parameter $\omega $\ of our model evolves from the
quintessence region to the $\Lambda CDM$ model region in the future $\omega
\rightarrow -1$.

\item All the energy conditions are satisfied throughout cosmic evolution,
except the SEC condition is violated, and the reason is due to cosmic
acceleration.

\item For the stability analysis, the background solution is stable against
the perturbation of the metric.

\item The jerk parameter is positive throughout the evolution of the
Universe. Thus, our model can be expected to adopt the behavior of another
DE model instead of the $\Lambda CDM$ model $j\neq 1$ at present $z=0$, but
our model is similar to the $\Lambda CDM$ model $j=1$ in the future $%
z\rightarrow -1$.

\item The statefinder parameters $\left\{ r,s\right\} $ evolve from the CG
region to the quintessence region at present, and later time to $\Lambda CDM$
point. As a result, our model behaves like the $\Lambda CDM$ model in the
future.
\end{itemize}

The obtained results are similar to several works that discuss the issue of
dark energy and cosmic acceleration in different contexts: $f\left(
R,T\right) $ gravity, $f\left( G\right) $ gravity, $f\left( T\right) $
gravity, etc. The only difference is the choice of a different background
for the study. In this reference \cite{ref58} Sharma et al. obtained similar
results for our model by studying the simplest non minimal matter-geometry
coupling in the framework of the $f\left( R,T\right) $ gravity with power
law expansion of the scale factor. It is noticeable that such forms of the
scale factor produce a constant deceleration parameter \cite{ref59, ref60,
ref61}, while in the present work we chose the deceleration parameter as a
linear function of the Hubble parameter which leads to the production of the
deceleration parameter varies with cosmic time, such as \cite{ref41, ref62}.

\section*{Acknowledgments}
We are very much grateful to the honorary referee and the
editor for the illuminating suggestions that have significantly improved our
work in terms of research quality and presentation.\newline

\textbf{Data availability} There are no new data associated with this article%
\newline

\textbf{Declaration of competing interest} The authors declare that they
have no known competing financial interests or personal relationships that
could have appeared to influence the work reported in this paper.\newline

\end{document}